\documentclass[english,aps,showpacs]{revtex4}
\usepackage[T1]{fontenc}
\usepackage[latin9]{inputenc}
\usepackage{graphicx}
\usepackage{amssymb}

\makeatletter

\providecommand{\tabularnewline}{\\}

\usepackage{babel}
\makeatother

\begin{document}

\preprint{This line only printed with preprint option}

\title{Microscopic justification of the Equal Filling approximation}

\author{Sara Perez-Martin}

\email{Sara.Perez@uam.es}

\altaffiliation[Currently at ]{CEA/DAM Île-de-France DPTA/Service de Physique Nucléaire, Bruyères-le-Châtel,
F-91297 Arpajon cedex, France}

\affiliation{Departamento de Física Teórica C-XI, Facultad de Ciencias, Universidad
Autónoma de Madrid, 28049 Madrid, Spain}


\author{L.M. Robledo}

\email{Luis.Robledo@uam.es}

\affiliation{Departamento de Física Teórica C-XI, Facultad de Ciencias, Universidad
Autónoma de Madrid, 28049 Madrid, Spain}

\begin{abstract}
The Equal Filling Approximation, a procedure widely
used in mean field calculations to treat the dynamics of odd nuclei
in a time reversal invariant way, is justified as the consequence
of a variational principle over an average energy functional. The
ideas of Statistical Quantum Mechanics are employed in the justification.
As an illustration of the method, the ground and lowest lying states
of some octupole deformed Radium isotopes are computed.
\end{abstract}

\pacs{21.60.Jz, 21.60.-n, 21.60.-k}

\maketitle

\section{introduction}

The Hartree-Fock-Bogoliubov (HFB) approximation is the cornerstone
of the microscopic description of the atomic nucleus as it encompasses
in the same approximation the concept of mean field orbits needed
to understand the extra stability associated with magic numbers as
well as the concept of pairing correlations needed to understand,
among others, why the ground state of all even-even nuclei always
have the $J^{\pi}=0^{+}$ quantum numbers. The HFB approximation has
widely been used over the last decades with great success both to
describe known properties of nuclei and to predict the properties
of unknown or yet to be experimentally studied nuclei \cite{Bender.05,RS.80}.

Another nice feature of the mean field approximation is that it allows
for the {}``spontaneous symmetry breaking'' (SSB) mechanism in which
the approximate (mean field) solution of the problem breaks the underlying
symmetries of the nuclear Hamiltonian. By the SSB mechanism many correlations
can be incorporated into the mean field wave function while maintaining
the simplicity of the mean field description. A minor (mainly practical)
drawback of the SSB mechanism is that the breaking of symmetries naturally
leads to the appearance of {}``full matrices'' in the numerical
implementation of the method. Those {}``full matrices'' are a direct
consequence of the mixing of quantum numbers that otherwise could
be used to bestow a block structure to the matrices considered. The
increase of the effective size of the matrices leads to an increase
in the number of operation needed to accomplish the numerical implementation
of the method and therefore leads to an increase in the computational
cost of the problem. Depending on the type of problem and the symmetry
broken, the computational cost can increase so dramatically as to
prevent the large scale calculations needed to describe stellar nucleosynthesis
or the stability of superheavy nuclei, just to mention a couple of
physical situations of interest nowadays. A typical case of a computationally demanding
application is the study of fission barriers allowing for time reversal
symmetry breaking, which is characteristic of high spin states or
odd mass nuclei which have to be treated in the framework of the standard
blocking method. In the treatment of odd mass systems we have an additional
source of computational complexity and it is the self-consistent character
of the HFB equations, which does not grant that blocking the quasiparticle
with lowest excitation energy will yield the lowest energy self-consistent
solution. As a consequence, several blocking possibilities have to
be considered multiplying the computing time by the number of the
possibilities considered (typically three or four times for each possible
$J_{z}$ and parity value).

On the other hand, the description of odd-A nuclei has started
to receive the attention it deserves (three quarters of all accessible
nuclei are odd-A ones) as the typical odd-even effects, which are not
well understood, are intimately related to pairing properties and
can also serve as more stringent guidelines to the development of
new energy functionals (see for instance \cite{Duguet.01,Zalewski.08}
for recent discussions on this issue). As a consequence, any attempt
to reduce the computational cost in the theoretical evaluation of
odd-A nuclei properties is very important as it will eventually 
allow for a systematic and routinely evaluation of the required
properties all over the periodic table.

A way to reduce the computational cost of mean field calculations
when dealing with odd mass nuclei is to try to keep time reversal
symmetry facilitating in this way the imposition of axial symmetry.
To keep time reversal symmetry when dealing with odd mass nuclei one
is forced to adopt phenomenological approaches in which the unpaired
nucleon is treated in an equal footing with its time reversed companion.
From a practical point of view this phenomenological approach amounts
to look at the unpaired nucleon as siting half in a given orbital 
and the other half in the time reversed partner (in the case
of preserving spherical symmetry where the orbitals have the $2j+1$
degeneracy the unpaired nucleon is distributed among all possible
angular momentum projections $m=-j,\ldots,j$ with equal probability
$1/(2j+1)$). The above procedure is usually referred in the literature
as the Equal (or uniform) Filling Approximation (EFA) and has been used
quite often in the description of odd nuclei at the mean field level
and with different interactions - see Refs. \cite{Berger.03,Sara.05,Baldini.06,Bonneau.07}
for recent applications of the method. This procedure is used
because it is considered as an {}``intuitive'' and
{}``reasonable'' approach but it is phenomenological in character and 
lacks a solid foundation as there is no product wave function that can
reproduce the density matrix and pairing tensor of the EFA. 
The purpose of this paper is to show that the Equal Filling Approximation 
(EFA) can be described in terms of a mixed state (in the sense of quantum 
statistical mechanics) density operator and the equations to be solved 
are a direct consequence of the variational principle over the energy of 
such mixed state. As a consequence of this microscopic justification 
it is now possible to introduce numerical procedures like the gradient 
method to solve the EFA equations facilitating enormously the procedure specially
in the case of many constraints. Another consequence is that now other
methods beyond mean field like the calculation of collective masses
or the Generator Coordinate Method itself can be consistently implemented
in the EFA framework.

Obviously, the EFA is an approximation to the correct treatment of odd-A nuclei
in the context of the mean field (the blocking procedure, see \cite{RS.80} for
a detailed explanation). To discuss the possible differences between them it is convenient
to look at the odd-A systems as made of an even-even core plus an unpaired nucleon
(or quasiparticle).
The interaction between the unpaired nucleon and the even-even core will induce
polarization effects in the core of three types 
\cite{Duguet.01,Zalewski.08,Rutz.98,Rutz.99,Xu.99}; namely the mass polarization,
the deformation polarization and the spin polarization effects. The only one
associated to the breaking of time reversal invariance is the spin polarization and it
is obvious that it is not included in the EFA. The other two effects (mass and shape
polarizations) are obviously included in the EFA to some extent but 
it is not easy to estimate the
possible impact of spin polarization effects on them. It is clear that a deeper
understanding of the interrelationship between the three effects is needed and
hopefully the present justification of the EFA will help to clarify the issue.

\section{The Equal Filling Approximation}

In the standard HFB method \cite{RS.80} quasi-particle operators
$\beta_{\mu}^{+}$ are introduced as linear combinations of the creation
and annihilation single particle operators corresponding to an arbitrarily
chosen (usually a harmonic oscillator) basis\begin{equation}
\beta_{\mu}^{+}=\sum_{m}U_{m\mu}c_{m}^{+}+V_{m\mu}c_{m}.\label{eq:Bogoliubov}\end{equation}
The HFB ground state wave function is defined by the condition of
being the vacuum of all the quasi-particle annihilation operators,
that is $\beta_{\mu}|\phi\rangle=0$. A more concise definition is
given by $|\phi\rangle=\prod_{\mu}\beta_{\mu}|0\rangle$ where the
index $\mu$ run over all the quasi-particle annihilation operators
that do not annihilate the true vacuum $|0\rangle$. The previous
results will describe the ground state of an even-even nucleus as
it can be shown that a paired HFB wave function is a linear combination
of product wave functions with an even number of particles. On the
other hand, odd-particle systems are handled by the {}``blocked''
HFB wave functions\begin{equation}
|\tilde{\phi}\rangle_{\mu_{B}}=\beta_{\mu_{B}}^{+}|\phi\rangle\label{eq:}\end{equation}
where $\mu_{B}$ is any of the quasi-particle indexes compatible with
the symmetries of the odd-particle system as, for instance, the $K$
quantum number (eigenvalue of $J_z$) or the parity. As the {}``blocked'' HFB wave function
is a product of quasi-particle operators, Wick's theorem applies and
the energy is given in the usual way in terms of the {}``blocked''
normal and abnormal densities $\rho^{(\mu_{B})}$ and $\kappa^{(\mu_{B})}$
\begin{equation}
E^{(\mu_{B})}=\mathrm{Tr}[t\rho^{(\mu_{B})}]+
\frac{1}{2}\mathrm{Tr}[\Gamma^{(\mu_{B})}\rho^{(\mu_{B})}]-
\frac{1}{2}\mathrm{Tr}[\Delta^{(\mu_{B})}\kappa^{(\mu_{B})*}]
\label{eq:}
\end{equation}
The normal and abnormal densities are given by
\begin{equation}
\rho_{kk'}^{(\mu_{B})}=
\langle\phi|\beta_{\mu_{B}}c_{k'}^{+}c_{k}
\beta_{\mu_{B}}^{+}|\phi\rangle=\left(V^{*}V^{T}\right)_{kk'}+
\left(U_{k'\mu_{B}}^{*}U_{k\mu_{B}}-V_{k'\mu_{B}}V_{k\mu_{B}}^{*}\right)
\label{eq:}
\end{equation}
and
\begin{equation}
\kappa_{kk'}^{(\mu_{B})}=
\langle\phi|\beta_{\mu_{B}}c_{k'}c_{k}\beta_{\mu_{B}}^{+}|\phi\rangle=
\left(V^{*}U^{T}\right)_{kk'}+
\left(U_{k\mu_{B}}V_{k'\mu_{B}}^{*}-
U_{k'\mu_{B}}V_{k\mu_{B}}^{*}\right)\label{eq:}
\end{equation}
These two matrices obviously violate time-reversal invariance. As
a consequence, the HF field 
\begin{equation}
\Gamma_{ll'}^{(\mu_{B})}=\sum_{qq'}\overline{v}_{lq'l'q}\rho_{qq'}^{(\mu_{B})}\label{g}
\end{equation}
as well as the pairing field
\begin{equation}
\Delta_{ll'}^{(\mu_{B})}=
\frac{1}{2}\sum_{qq'}\overline{v}_{ll'qq'}\kappa_{qq'}^{(\mu_{B})}\label{d}
\end{equation}
both violate time-reversal invariance making the numerical calculation
much more computationally expensive to carry out. A way to preserve
time-reversal invariance is to use the Equal Filling Approximation
(EFA) that amounts to use the {}``average'' density 
\begin{equation}
\rho_{kk'}^{EFA}=
\left(V^{*}V^{T}\right)_{kk'}+
\frac{1}{2}\left(U_{k'\mu_{B}}U_{k\mu_{B}}^{*}-
V_{k'\mu_{B}}^{*}V_{k\mu_{B}}+U_{k'\overline{\mu}_{B}}
U_{k\overline{\mu}_{B}}^{*}-
V_{k'\overline{\mu}_{B}}^{*}V_{k\overline{\mu}_{B}}\right)\label{eq:ROEFA}
\end{equation}
and the {}``average'' pairing tensor 
\begin{equation}
\kappa_{kk'}^{EFA}=\left(V^{*}U^{T}\right)_{kk'}+\frac{1}{2}
\left(U_{k\mu_{B}}V_{k'\mu_{B}}^{*}-U_{k'\mu_{B}}V_{k\mu_{B}}^{*}+
U_{k\overline{\mu}_{B}}V_{k'\overline{\mu}_{B}}^{*}-
U_{k'\overline{\mu}_{B}}V_{k\overline{\mu}_{B}}^{*}\right)\label{eq:KAEFA}
\end{equation}
that now preserve time reversal invariance as both expressions involve
and average with equal weights of the blocked level $\mu_{B}$ and
its time reversed and degenerate partner $\overline{\mu}_{B}$ (see below for 
higher order degeneracy). Intuitively, the
above densities should correspond to an occupancy of $1/2$ for the
states $\mu_{B}$ and $\overline{\mu}_{B}$. In the next step of the
EFA framework, it is assumed without proof that the energy is given
by the standard HFB expression but using $\rho^{EFA}$ and $\kappa^{EFA}$
instead of the corresponding densities, i.e. 
\begin{equation}
E_{EFA}=\mathrm{Tr}[t\rho^{EFA}]+\frac{1}{2}\mathrm{Tr}[\Gamma^{EFA}
\rho^{EFA}]-\frac{1}{2}\mathrm{Tr}[\Delta^{EFA}\kappa^{EFA*}]
\label{eq:E_EFA}
\end{equation}
Finally, it is assumed that the $U$ and $V$ amplitudes of the Bogoliubov
transformation are given as the solution of the standard HFB equation
\begin{equation}
\left(\begin{array}{cc}
h^{EFA} & \Delta^{EFA}\\
-\Delta^{EFA*} & -h^{EFA*}\end{array}\right)\left(\begin{array}{cc}
U & V^{*}\\
V & U^{*}\end{array}\right)=\left(\begin{array}{cc}
U & V^{*}\\
V & U^{*}\end{array}\right)\left(\begin{array}{cc}
E & 0\\
0 & -E\end{array}\right)\label{eq:HFBEQ_EFA}\end{equation}
where $E$ are the quasi-particle energies. To our knowledge, the
two previous assumptions of the EFA, namely that the energy is given
by Eq. (\ref{eq:E_EFA}) and that the $U$ and $V$ amplitudes are
given by Eq. (\ref{eq:HFBEQ_EFA}), lacked a foundation and were just
considered as a plausible quantity (the energy) and equation. Here
we will show that both assumptions are well founded in terms of standard
quantum mechanic procedures and therefore we are giving more credit
to the approximation.

\subsection{Justification of the EFA expression for the energy}

In the standard HFB theory the density matrix and pairing tensor are
the components of a bipartite generalized density matrix\begin{equation}
\mathcal{R}=\left(\begin{array}{cc}
\rho & \kappa\\
-\kappa^{*} & 1-\rho^{*}\end{array}\right)=\left(\begin{array}{cc}
U & V^{*}\\
V & U^{*}\end{array}\right)\left(\begin{array}{cc}
0 & 0\\
0 & 1\end{array}\right)\left(\begin{array}{cc}
U^{+} & V^{+}\\
V^{T} & U^{T}\end{array}\right)=W\mathbb{R}W^{+}\label{eq:}\end{equation}
where the generalized quasi-particle density matrix\begin{equation}
\mathbb{R}_{\nu\mu}=\left(\begin{array}{cc}
\left\langle \phi\right|\beta_{\mu}^{\dagger}\beta_{\nu}\left|\phi\right\rangle  & \left\langle \phi\right|\beta_{\mu}\beta_{\nu}\left|\phi\right\rangle \\
\left\langle \phi\right|\beta_{\mu}^{\dagger}\beta_{\nu}^{\dagger}\left|\phi\right\rangle  & \left\langle \phi\right|\beta_{\mu}\beta_{\nu}^{\dagger}\left|\phi\right\rangle \end{array}\right)=\left(\begin{array}{cc}
0 & 0\\
0 & 1\end{array}\right)\label{eq:}\end{equation}
and the Bogoliubov super-matrix\begin{equation}
W=\left(\begin{array}{cc}
U & V^{*}\\
V & U^{*}\end{array}\right)\label{eq:}\end{equation}
have been introduced. In the EFA case we can also introduce a generalized
density matrix\begin{equation}
\mathcal{R}^{EFA}=\left(\begin{array}{cc}
\rho^{EFA} & \kappa^{EFA}\\
-\kappa^{EFA*} & 1-\rho^{EFA*}\end{array}\right)\label{eq:BIGRO_EFA}\end{equation}
that again can be written as\begin{equation}
\mathcal{R}^{EFA}=W\mathbb{R}^{EFA}W^{+}\label{eq:ROEFA1}\end{equation}
with\begin{equation}
\mathbb{R}_{\nu\mu}^{EFA}=\left(\begin{array}{cc}
f_{\mu} & 0\\
0 & 1-f_{\mu}\end{array}\right)\label{eq:ROEFAQpart}\end{equation}
and the $f_{\mu}$ is given by\begin{equation}
f_{\mu}=\left\{ \begin{array}{ccl}
\frac{1}{2} &  & \mu=\mu_{B}\;\textrm{{or}}\;\overline{\mu}_{B}\\
0 &  & \textrm{otherwise}\end{array}\right.\label{eq:Statfac}\end{equation}
The above result immediately remind us of the Finite Temperature HFB
formalism \cite{Blaizot.86} where the quasi-particle density matrix has exactly the
same form as above but with the statistical occupancies \begin{equation}
f_{\mu}=\frac{1}{e^{\beta E_{\mu}}+1}\label{eq:}\end{equation}
Therefore, the EFA can be viewed as a Finite Temperature HFB formalism
with the statistical factors of Eq. (\ref{eq:Statfac}). The finite
temperature formalism is nothing but a quantum mechanics statistical
formalism where instead of pure states a statistical admixture of
them is considered weighted with given probabilities. In the finite
temperature formalism the probabilities are obtained according to
the statistical {}``ensemble'' considered but in the EFA they are
just fixed by the requirements of the approximation. For this reason
we will not use in the following the language of finite temperature
but instead the one of statistical quantum mechanics. The two relevant
concepts in statistical quantum mechanics are the one of the {}``density
matrix operator'' and the other is the concept of trace. In the present
context the trace is taken over the whole Fock space in such a way
that given a set of quasi-particle creation and annihilation operators
$\beta_{\mu}^{+}$ and $\beta_{\mu}$ and the corresponding vacuum
$|\phi\rangle$ (see Eq. (\ref{eq:Bogoliubov}) ) we have the following
expression for the trace of an arbitrary operator 
\begin{equation}
\textrm{Tr}[\hat{O}]=\left\langle \phi\right|\hat{O}\left|\phi\right\rangle +
\sum_{\mu}\left\langle \phi\right|\beta_{\mu}\hat{O}
\beta_{\mu}^{\dagger}\left|\phi\right\rangle +
\frac{1}{2!}\sum_{\nu\mu}\left\langle \phi\right|
\beta_{\mu}\beta_{\nu}\hat{O}\beta_{\nu}^{\dagger}
\beta_{\mu}^{\dagger}\left|\phi\right\rangle \ldots\label{eq:}\end{equation}
On the other hand, the density operator $\hat{\mathcal{D}}$ can be
chosen in such a way that $\hat{\mathcal{D}}|\phi\rangle=|\phi\rangle$
and $\hat{\mathcal{D}}\beta_{\mu}^{\dagger}=p_{\mu}\beta_{\mu}^{\dagger}\hat{\mathcal{D}}$
where $p_{\mu}$ is the probability of the one-quasi-particle excitation
$\beta_{\mu}^{+}|\phi\rangle$. In this formalism the statistical
mean value of an operator is given by \begin{equation}
\langle\hat{O}\rangle_{S}=\frac{\textrm{Tr}[\hat{O}\hat{\mathcal{D}}]}
{\textrm{Tr}[\hat{\mathcal{D}}]}=
\frac{1}{Z}\left(\left\langle \phi\right|\hat{O}\left|\phi\right\rangle +
\sum_{\mu}p_{\mu}\left\langle \phi\right|\beta_{\mu}\hat{O}\beta_{\mu}^{\dagger}\left|\phi\right\rangle +
\frac{1}{2!}\sum_{\nu\mu}p_{\mu}p_{\nu}\left\langle \phi\right|\beta_{\mu}\beta_{\nu}\hat{O}
\beta_{\nu}^{\dagger}\beta_{\mu}^{\dagger}\left|\phi\right\rangle \ldots\right)\label{eq:MVS}
\end{equation}
with
\begin{equation}
\textrm{Tr}[\hat{\mathcal{D}}]=Z=1+\sum_{\mu}p_{\mu}+\sum_{\nu<\mu}p_{\mu}p_{\nu}\ldots=\prod_{\mu}(1+p_{\mu})\label{eq:Z}
\end{equation}
It is also easy to show that 
\begin{equation}
\langle\beta_{\rho}\beta_{\sigma}\rangle_{S}=\langle\beta_{\rho}^{+}\beta_{\sigma}^{+}\rangle_{S}=0\label{eq:}
\end{equation}
and also
\begin{equation}
\langle\beta_{\rho}^{+}\beta_{\sigma}\rangle_{S}=\delta_{\rho\sigma}\frac{p_{\sigma}}{1+p_{\sigma}}=\delta_{\rho\sigma}f_{\sigma};\;\langle\beta_{\rho}\beta_{\sigma}^{+}\rangle_{S}=\delta_{\rho\sigma}\left(1-\frac{p_{\sigma}}{1+p_{\sigma}}\right)=\delta_{\rho\sigma}(1-f_{\sigma})\label{eq:}\end{equation}
and therefore we recover the EFA's density matrix of Eq. (\ref{eq:BIGRO_EFA})
by using the above formalism with 
\begin{equation}
p_{\mu}=\left\{ \begin{array}{ccc}
1 &  & \mu=\mu_{B}\;\textrm{{or}}\;\overline{\mu}_{B}\\
0 &  & \textrm{otherwise}\end{array}\right.\label{eq:P_EFA}
\end{equation}
Here we are implicitly assuming that the single particle levels are
doubly degenerate (Kramer's degeneracy) but in those cases where spherical
symmetry is preserved in the mean field procedure we will have to
populate with the same probability all the states with different $m=-j,\ldots,j$
(third component of the angular momentum) for a given orbital labeled
with the $j$ quantum number. The formalism being developed here apply
equally well in this case and the reader is referred to Appendix A
for technical details in the spherical case.

Thanks to the existence of a statistical Wick's theorem (see for instance
the proof given by Gaudin \cite{Gaudin.60} or \cite{SPM.07} for
a more recent account) it is possible to compute any statistical mean
value of a product of creation and annihilation operators in terms
of the corresponding contractions and therefore it is possible to
express the statistical mean value of the energy $\langle\hat{H}\rangle_{S}=\textrm{Tr}[\hat{H}\hat{\mathcal{D}}]/\textrm{Tr}[\hat{\mathcal{D}}]$
by using the standard expression \begin{equation}
\langle\hat{H}\rangle_{S}=\mathrm{Tr}[t\rho]+\frac{1}{2}\mathrm{Tr}[\Gamma\rho]-\frac{1}{2}\mathrm{Tr}[\Delta\kappa^{*}]\label{eq:}\end{equation}
where the density matrix and pairing tensor are given by the contractions\begin{equation}
\rho_{kk'}=\frac{\textrm{Tr}[c_{k'}^{+}c_{k}\hat{\mathcal{D}}]}
{\textrm{Tr}[\hat{\mathcal{D}}]};\:\:
\kappa_{kk'}=\frac{\textrm{Tr}[c_{k'}c_{k}\hat{\mathcal{D}}]}
{\textrm{Tr}[\hat{\mathcal{D}}]}\label{eq:}\end{equation}
Applying this result to the EFA case, we conclude that the energy
of Eq. (\ref{eq:E_EFA}) can be written as $E_{EFA}=\textrm{Tr}[\hat{H}\hat{\mathcal{D}}^{EFA}]/\textrm{Tr}[\hat{\mathcal{D}}^{EFA}]$.
This result justifies the, otherwise ad-hoc, expression of the EFA
energy and gives a physical interpretation to it as the statistical
mean value of the Hamiltonian taken with the EFA density operator.
The EFA energy $E_{EFA}$ can also be written in a more transparent
way by using Eq. (\ref{eq:MVS}) together with Eq. (\ref{eq:P_EFA})
as\begin{equation}
E_{EFA}=\frac{1}{4}\left(\left\langle \phi\right|\hat{H}\left|\phi\right\rangle +\left\langle \phi\right|\beta_{\mu_{B}}\hat{H}\beta_{\mu_{B}}^{\dagger}\left|\phi\right\rangle +\left\langle \phi\right|\beta_{\overline{\mu}_{B}}\hat{H}\beta_{\overline{\mu}_{B}}^{\dagger}\left|\phi\right\rangle +\left\langle \phi\right|\beta_{\mu_{B}}\beta_{\overline{\mu}_{B}}\hat{H}\beta_{\overline{\mu}_{B}}^{\dagger}\beta_{\mu_{B}}^{\dagger}\left|\phi\right\rangle \right)\label{eq:}\end{equation}
showing that it is simply an average with
equal weights of the energy of the reference even-even wave function
$|\phi\rangle$, the energies of one quasi-particle excitations with
quantum numbers $\mu_{B}$ and $\overline{\mu}_{B}$ and the energy
of the two quasi-particle excitation with the same quantum numbers.
This result, which was to the knowledge of the authors previously
unknown\footnote{In Ref. \cite{Duguet.01} it is mentioned that
the EFA  in the zero pairing limit can be described
by the density operator of a mixed state but no further development of the idea was pursued there.}, 
is very illustrative of the nature of the EFA as a statistical
theory. The same kind of arguments can be applied to compute mean
values of any kind of operators in the EFA framework. A curious result
that can be easily derived is that the EFA mean values of any one-body
operator, which according to the general result can be written as
\begin{equation}
\langle\hat{O}\rangle_{EFA}=\frac{1}{4}\left(\left\langle \phi\right|\hat{O}\left|\phi\right\rangle +\left\langle \phi\right|\alpha_{\mu_{B}}\hat{O}\alpha_{\mu_{B}}^{\dagger}\left|\phi\right\rangle +\left\langle \phi\right|\alpha_{\overline{\mu}_{B}}\hat{O}\alpha_{\overline{\mu}_{B}}^{\dagger}\left|\phi\right\rangle +\left\langle \phi\right|\alpha_{\mu_{B}}\alpha_{\overline{\mu}_{B}}\hat{O}\alpha_{\overline{\mu}_{B}}^{\dagger}\alpha_{\mu_{B}}^{\dagger}\left|\phi\right\rangle \right),\label{eq:}\end{equation}
can also be written in a more compact way as\begin{equation}
\langle\hat{O}\rangle_{EFA}=\frac{1}{2}\left(\left\langle \phi\right|\alpha_{\mu_{B}}\hat{O}\alpha_{\mu_{B}}^{\dagger}\left|\phi\right\rangle +\left\langle \phi\right|\alpha_{\overline{\mu}_{B}}\hat{O}\alpha_{\overline{\mu}_{B}}^{\dagger}\left|\phi\right\rangle \right)=\frac{1}{2}\left(\left\langle \phi\right|\hat{O}\left|\phi\right\rangle +\left\langle \phi\right|\alpha_{\mu_{B}}\alpha_{\overline{\mu}_{B}}\hat{O}\alpha_{\overline{\mu}_{B}}^{\dagger}\alpha_{\mu_{B}}^{\dagger}\left|\phi\right\rangle \right)\label{eq:}\end{equation}
This allows us to write the density matrix and pairing tensor as an
average over one quasi-particle excitations\begin{equation}
\rho_{kk'}^{EFA}=\frac{1}{2}\left(\langle\phi|\alpha_{\mu_{B}}c_{k'}^{+}c_{k}\alpha_{\mu_{B}}^{+}|\phi\rangle+\langle\phi|\alpha_{\bar{\mu}_{B}}c_{k'}^{+}c_{k}\alpha_{\bar{\mu}_{B}}^{+}|\phi\rangle\right)\label{eq:}\end{equation}
and \begin{equation}
\kappa_{kk'}^{EFA}=\frac{1}{2}\left(\langle\phi|\alpha_{\mu_{B}}c_{k'}c_{k}\alpha_{\mu_{B}}^{+}|\phi\rangle+\langle\phi|\alpha_{\bar{\mu}_{B}}c_{k'}c_{k}\alpha_{\bar{\mu}_{B}}^{+}|\phi\rangle\right)\label{eq:}\end{equation}
which is a very intuitive result according to the expressions of Eqs.
(\ref{eq:ROEFA}) and (\ref{eq:KAEFA}). This result, however, does
by no means imply that the energy, which is the average of a two-body
operator, could be written as $\frac{1}{2}\left(\langle\phi|\alpha_{\mu_{B}}H\alpha_{\mu_{B}}^{+}|\phi\rangle+\langle\phi|\alpha_{\bar{\mu}_{B}}H\alpha_{\bar{\mu}_{B}}^{+}|\phi\rangle\right)$.

\subsection{Variational derivation of the EFA-HFB equation}

Another interesting feature of the results obtained so far is that
the variational principle can be applied now to the $E_{EFA}$ energy.
As it will be shown below the variational principle leads naturally
to the HFB equation of the EFA mentioned in Eq (\ref{eq:HFBEQ_EFA})
justifying thereof its use to determine the coefficients of the Bogoliubov
transformation. This result is also advantageous from a practical
point of view as the variational origin of the HFB-EFA equations allows
the use of {}``gradient-like'' methods to solve it and also makes
the treatment of constraints much easier.

The first step in the application of the variational principle is
to establish the variational space by defining the most general Bogoliubov
transformation. This is a common procedure that the interested reader
can consult in the standard literature \cite{RS.80,RREF.84} and
we will give here only the most relevant formulas just to establish
the notation. Given a reference HFB wave function $|\phi\rangle$
the most general HFB wave function $|\phi(\mathbb{Z})\rangle$ not
orthogonal to it is given by $|\phi(\mathbb{Z})\rangle=\exp(i\hat{Z})|\phi\rangle$
where $\hat{Z}$ is an hermitian (to preserve the unitarity of the
transformation) one-body operator $\hat{Z}=\frac{1}{2}\sum_{\mu\nu}\mathbb{Z}_{\mu\nu}\alpha_{\mu}^{+}\alpha_{\nu}$
which is here written in terms of the generalized quasi-particle operators
$\alpha_{\mu}=(\beta_{1},\ldots,\beta_{N},\beta_{1}^{+},\ldots,\beta_{N}^{+})$,
its hermitian conjugate $\alpha_{\mu}^{+}$ and the bipartite hermitian
matrix \cite{RREF.84}\begin{equation}
\mathbb{Z}=\left(\begin{array}{cc}
Z^{11} & Z^{20}\\
-Z^{20*} & -Z^{11*}\end{array}\right)\label{eq:}
\end{equation}
where $Z_{mn}^{11}$ ($m,n=1,\ldots,N$) is an hermitian matrix ($N^{2}$ free parameters,
complex $Z_{mn}^{11}$ with $m>n$ plus real $Z_{mm}^{11}$ ) whereas
$Z^{20}$ is skew-symmetric ($N^{2}-N$ free parameters, complex $Z_{mn}^{20}$
with $m>n$ ). The matrix elements of $Z_{mn}^{11}$ with $m>n$ plus
$Z_{mm}^{11}$ and those of $Z_{mn}^{20}$ with $m>n$ constitute
the complex variational parameters of our model. The complex variational
parameters will also be denoted by the vector $z_{\rho}$ of dimension
$2N^{2}-N$ (see appendix B for details). The coefficients of the
Bogoliubov transformation of the quasi-particle operators associated
to $|\phi(\mathbb{Z})\rangle$ are given by the matrix $W(\mathbb{Z})=W(0)\exp(i\mathbb{Z})$
which is written in terms of the exponential of $\mathbb{Z}$ and the Bogoliubov
transformation coefficients $W(0)$ of the reference HFB wave function
$|\phi\rangle$. To determine the dependence of the generalized density
with the variational parameters we will use Eq. (\ref{eq:ROEFA1})
to write
\begin{equation}
\mathcal{R}^{EFA}(\mathbb{Z})=W(\mathbb{Z})\mathbb{R}^{EFA}W^{+}(\mathbb{Z})\label{eq:}
\end{equation}
where we have kept $\mathbb{R}^{EFA}$ fixed as in Eq. (\ref{eq:ROEFAQpart}),
according to its definition. For infinitesimal variational parameters
(i.e. infinitesimal $\mathbb{Z}$) we have
\begin{equation}
\mathcal{R}^{EFA}(\mathbb{Z})=\mathcal{R}^{EFA}(0)+iW(0)[\mathbb{Z},\mathbb{R}^{EFA}]W^{+}(0)+
O(\mathbb{Z}^{2})=\mathcal{R}^{EFA}(0)+i[\mathcal{Z},\mathcal{R}^{EFA}]+O(\mathbb{Z}^{2})\label{eq:DeltaRo}
\end{equation}
where $\mathcal{Z}=W(0)\mathbb{Z}W^{+}(0)$. Now to facilitate the
manipulation of different quantities we write the energy as 
\begin{equation}
E_{EFA}=\frac{1}{4}\mathrm{Tr_{2}}\left[(\mathcal{H}+\mathcal{T})\mathcal{S}\right]\label{eq:EEFA}
\end{equation}
in terms of the Hamiltonian matrix $\mathcal{H}=\left(\begin{array}{cc}
t+\Gamma & \Delta\\
-\Delta^{*} & -(t+\Gamma)^{*}\end{array}\right)$, the kinetic energy matrix $\mathcal{T}=\left(\begin{array}{cc}
t & 0\\
0 & -t^{*}\end{array}\right)$ and the matrix $\mathcal{S}=\mathcal{R}-\left(\begin{array}{cc}
0 & 0\\
0 & 1\end{array}\right)=\left(\begin{array}{cc}
\rho & \kappa\\
-\kappa^{*} & -\rho^{*}\end{array}\right)$ derived from the generalized density. 
The trace is taken over the
double size space where bipartite matrices are defined. In order to
arrive to the above expression we have made use of the properties
$\mathrm{Tr}[\Gamma_{2}\rho_{1}]=\mathrm{Tr}[\Gamma_{1}\rho_{2}]$
and $\mathrm{Tr}[\Delta_{2}\kappa_{1}^{*}]=\mathrm{Tr}[\Delta_{1}\kappa_{2}^{*}]^{*}$
where $\Gamma_{i}$ and $\Delta_{i}$ stand for the Hartree-Fock and
pairing fields computed with the density matrix $\rho_{i}$ and pairing
tensor $\kappa_{i}$ respectively. The two previous relations can
be written using bipartite matrices as 
\begin{equation}
\mathrm{Tr_{2}\left[(\mathcal{H}_{1}-\mathcal{T})\mathcal{S}_{2}\right]}=
Tr_{2}\left[(\mathcal{H}_{2}-\mathcal{T})\mathcal{S}_{1}\right]\label{eq:}
\end{equation}
This result allows to write the variation of the energy in a more
compact way, namely
\begin{equation}
\delta E_{EFA}=\frac{1}{4}\left(\mathrm{Tr_{2}}\left[
(\mathcal{H}+\mathcal{T})\delta\mathcal{S}\right]+
\mathrm{Tr_{2}}\left[(\delta\mathcal{H}-\mathcal{T})
\mathcal{S}\right]\right)=\frac{1}{2}
\mathrm{Tr_{2}}\left[\mathcal{H}\delta\mathcal{S}\right]=
\frac{i}{2}\mathrm{Tr_{2}}\left[[\mathcal{R},\mathcal{H}]
\mathcal{Z}\right]+O(\mathbb{Z}^{2})\label{eq:deEFA}
\end{equation}
where we have made use of Eq. (\ref{eq:DeltaRo}) and the fact that
$\delta\mathcal{S}=\delta\mathcal{R}$. The variational condition
$\delta E_{EFA}=0$ has to be handled with care as not all the parameters
of the bipartite matrix $\mathcal{Z}$ are variational parameters
but as it is shown in appendix B the variational condition can be
written as $[\mathcal{R},\mathcal{H}]=0$ which is the standard form
of the HFB equation. It has to be emphasized again that the previous
form of the HFB equation in the EFA was just an assumption and now
we are able to justify it in our framework by simply invoking the
variational principle. The HFB equation could be solved using iterative
methods as it is customary with the standard HFB equations but we
have found more convenient to take advantage of the variational origin
of the equation in order to use other methods to find its minimum,
like the gradient method. To implement the gradient method it is more
convenient to write the variation of the EFA energy in the {}``quasi-particle''
basis as 
\begin{equation}
\delta E_{EFA}=\frac{i}{2}
\mathrm{Tr_{2}}\left[[\mathbb{R},\mathbb{H}]
\mathbb{Z}\right]+O(\mathbb{Z}^{2})\label{eq:}
\end{equation}
with $\mathbb{H=}W^{+}(0)\mathcal{H}W(0)=\left(\begin{array}{cc}
H^{11} & H^{20}\\
-H^{20\:*} & -H^{11\:*}\end{array}\right)$. Using this form and the definitions of appendix B we can finally
write $\delta E_{EFA}=\sum_{\rho=1}^{2N^{2}-N}\left(g_{E}\right)_{\rho}z_{\rho}+O(z^{2})$
where $\left(g_{E}\right)_{\rho}$ are the components of the gradient
of the energy with respect to the variational parameters $z_{\rho}$.
In the spirit of the gradient method, by choosing $z_{\rho}=-\eta\left(g_{E}^{*}\right)_{\rho}$
we make sure we gain energy at least to first order in $z$ if the
scaling parameter (or step size) $\eta$ is always chosen to be positive.
The previous election implies that $\mathbb{Z}=i\eta[\mathbb{R},\mathbb{H}]$.
The step size $\eta$ is estimated in each iteration as to make second
order terms smaller enough compared with first order ones so that
the energy always decreases. Once the $\mathbb{Z}$ parameters have
been determined the wave function $W(\mathbb{Z})$ is computed by
evaluating the exponential $e^{i\mathbb{Z}}$ by means of a Padé approximation
to the exponential of the form $e^{x}=N_{pp}(x)/N_{pp}(-x)$ with
$N_{pp}(x)=\sum_{k=0}^{p}c_{k}^{(p)}x^{k}$ (that is, the polynomials
in both the numerator and denominator have the same degree; the coefficients
$c_{k}^{(p)}$ are determined by the standard recurrence relation
$c_{k}^{(p)}=c_{k-1}^{(p)}\frac{p+1-k}{(2p+1-k)k}$ ) that has the
nice feature of preserving the unitarity of the exponential when used
with anti-hermitian exponents as it is the case. Usually, a Padé approximation
of order $p=1$ suffices and in our numerical implementation we have taken
$e^{i\mathbb{Z}}\approx(\openone+\frac{i}{2}\mathbb{Z})(\openone-\frac{i}{2}\mathbb{Z})^{-1}$.

\subsection{Dealing with constraints}

In dealing with constraints, we face the numerical
problem of minimizing a given function of the variational parameters
$E_{EFA}(\mathbb{Z})$ subject to some constraints of the kind
\begin{equation}
\frac{\textrm{Tr}[\hat{Q}_{i}\hat{\mathcal{D}}]}{\textrm{Tr}[\hat{\mathcal{D}}]}=q_{i}\label{eq:}
\end{equation}
 where $\hat{Q}_{i}$ are one-body operators which can be usually
written as $\hat{Q}_{i}=\sum_{kk'}\left(Q_{i}\right)_{kk'}c_{k}^{+}c_{k'}$
(extending the results below to the case of operators not commuting
with the particle number one is straightforward). Introducing the
bipartite matrix $\mathcal{Q}_{i}=\left(\begin{array}{cc}
Q_{i} & 0\\
0 & -Q_{i}^{*}\end{array}\right)$ we can write 
\begin{equation}
q_{i}=\frac{1}{2}\mathrm{Tr}[\mathcal{Q}_{i}\mathcal{S}]\label{eq:qi}
\end{equation}
that yields, in analogy to Eq (\ref{eq:deEFA}), to the following expression
for the variation of $q_{i}$
\begin{equation}
\delta q_{i}=
\frac{i}{2}\mathrm{Tr_{2}}\left[[\mathcal{R},\mathcal{Q}_{i}]\mathcal{Z}\right]+
O(\mathbb{\mathcal{Z}}^{2})=\frac{i}{2}\mathrm{Tr_{2}}\left[[\mathbb{R},\mathbb{Q}_{i}]
\mathbb{Z}\right]+O(\mathbb{Z}^{2})\label{eq:dqi}
\end{equation}
To consider the constrained minimization procedure we will proceed
in the standard way by introducing Lagrange multipliers $\lambda_{i}$
and a new functional to be minimized, namely 
$E'_{EFA}(\mathbb{Z})=E{}_{EFA}(\mathbb{Z})-\sum_{i}\lambda_{i}q_{i}(\mathbb{Z})$.
The Lagrange multipliers are determined as to make the gradient of
$E'_{EFA}(\mathbb{Z})$ orthogonal to the ones of the $q_{i}(\mathbb{Z})$'s
that will be denoted by $\left(g_{q_{i}}\right)_{\rho}$. Taking into
account that the gradient of $E'_{EFA}(\mathbb{Z})$ is given by the
vector $\left(g_{E'}\right)_{\rho}=\left(g_{E}\right)_{\rho}-\sum_{j}\lambda_{j}\left(g_{q_{j}}\right)_{\rho}$
the orthogonality condition yields
\begin{equation}
\lambda_{i}=\sum_{j}S_{ij}^{-1}d_{j}\label{eq:}
\end{equation}
where
\begin{equation}
S_{ij}=\sum_{\rho}\left(g_{q_{i}}\right)_{\rho}^{*}\left(g_{q_{j}}\right)_{\rho}\label{eq:}
\end{equation}
and 
\begin{equation}
d_{i}=\sum_{\rho}\left(g_{E}\right)_{\rho}^{*}\left(g_{q_{j}}\right)_{\rho}\label{eq:}
\end{equation}
Using the explicit form of the corresponding gradients (see appendix
B) we can finally write the two quantities above as
\begin{equation}
S_{ij}=\frac{1}{4}\mathrm{Tr}_{2}[[\mathbb{Q}_{i},\mathbb{R}][\mathbb{Q}_{j},\mathbb{R}]]\label{eq:Sij}
\end{equation}
and
\begin{equation}
d_{i}=\frac{1}{4}\mathrm{Tr}_{2}[[\mathbb{Q}_{i},\mathbb{R}][\mathbb{H},\mathbb{R}]]\label{eq:di}
\end{equation}
It is also convenient to establish a procedure to readjust the constraints
as it is customary that in the iterative process their values $q_{i}$
slightly depart from the desired ones $q_{i}^{(0)}$, that is $q_{i}=q_{i}^{(0)}+\delta q_{i}$.
In the context of the gradient method such a procedure is elemental
and we only have to replace the chemical potentials $\lambda_{j}$
by $\lambda_{j}+\delta\lambda_{j}$ and impose that the variation
of the values of the constraints $\delta q_{i}$ given by Eq. (\ref{eq:dqi})
yields the desired value. The result is $\eta\delta\lambda_{j}=\sum_{i}S_{ij}^{-1}\delta q_{i}$
where the matrix elements $S_{ij}$ are the same as in Eq. (\ref{eq:Sij}).

\subsection{Density dependent interactions}

For density dependent interactions, like the Gogny force \cite{Decharge.80}
used in the next section to illustrate the whole procedure, we have
to define the explicit form of the density dependent part of the interaction
for statistical averages. It seems natural that, if for a pure state
(an HFB mean field wave function in this case) the DD part of the
interaction is a function of the density of the pure state, then for
a statistical average the DD should be the same function but of the
density of the statistical average; that is, a function of 
\begin{equation}
\rho(\vec{R})=\frac{\textrm{Tr}[\hat{\rho}(\vec{R})\hat{\mathcal{D}}]}
{\textrm{Tr}[\hat{\mathcal{D}}]}=\frac{1}{Z}
\left(\left\langle \phi\right|\hat{\rho}(\vec{R})\left|\phi\right\rangle 
+\sum_{\mu}p_{\mu}\left\langle \phi\right|\beta_{\mu}
\hat{\rho}(\vec{R})\beta_{\mu}^{\dagger}\left|\phi\right\rangle +
\frac{1}{2!}\sum_{\nu\mu}p_{\mu}p_{\nu}\left\langle 
\phi\right|\beta_{\mu}\beta_{\nu}\hat{\rho}(\vec{R})
\beta_{\nu}^{\dagger}\beta_{\mu}^{\dagger}\left|\phi\right\rangle 
\ldots\right)\label{eq:}
\end{equation}
(see Eq. (\ref{eq:Z}) for the definition of $Z$). This prescription
has been the one used in previous calculations with the Gogny force
at finite temperature \cite{Temp1,Temp2} as well as by other authors
with other density dependent interactions like several variants of
the Skyrme one \cite{Temp3}. This prescription has the right limit
when the probabilities go to zero (pure state) and also produces consistent
results when the one-quasi-particle energies are computed as partial
derivatives of the energy with respect to the probabilities: when
the above prescription is used the expression for the one-quasi-particle
energies includes the rearrangement term present in all the HF or
HFB calculations with density dependent interactions \cite{Decharge.80}.
It is obvious that for a consistent treatment of the problem, the
variation of the energy with respect to the variational parameters
has to take into account also that the Hamiltonian depends upon them
via the DD term and the corresponding rearrangement terms have to
be considered (see \cite{Decharge.80,GradC} for details). To summarize
this section, in the EFA case we use the density
\begin{equation}
\rho^{EFA}(\vec{R})=
\frac{\textrm{Tr}[\hat{\rho}(\vec{R})\hat{\mathcal{D}}^{EFA}]}
{\textrm{Tr}[\hat{\mathcal{D}}^{EFA}]}=
\frac{1}{4}\left(\left\langle \phi\right|
\hat{\rho}(\vec{R})\left|\phi\right\rangle +
\left\langle \phi\right|\beta_{\mu_{B}}
\hat{\rho}(\vec{R})\beta_{\mu_{B}}^{\dagger}
\left|\phi\right\rangle +\left\langle 
\phi\right|\beta_{\bar{\mu}_{B}}
\hat{\rho}(\vec{R})\beta_{\bar{\mu}_{B}}^{\dagger}
\left|\phi\right\rangle +
\left\langle \phi\right|\beta_{\mu_{B}}
\beta_{\bar{\mu}_{B}}\hat{\rho}(\vec{R})
\beta_{\bar{\mu}_{B}}^{\dagger}
\beta_{\mu_{B}}^{\dagger}
\left|\phi\right\rangle \right)\label{eq:}
\end{equation}
for the density dependent part of the Gogny interaction.

\section{Results}

To show an example of the proposed method we have computed the spectrum
of several odd-A isotopes of the Radium in the range of $A$ between
221 and 231. As it is well known, some isotopes of Ra are known to
display octupole deformation \cite{Butler.96} in their ground state
and therefore, in order to study their spectrum, we will carry out
calculations constraining the octupole moment to locate the different
minima and to check their depths as they are relevant for the stability
of the configuration against octupole fluctuations. We will limit
the calculation to axially symmetric (but reflection symmetry breaking)
configurations and therefore each of the blocked levels will be characterized
(and labeled) by its $J_{z}$ value (but not parity). The calculations
are performed in the framework of the EFA with the finite range and
effective interaction of Gogny \cite{Decharge.80}. As it is customary
in all the mean field calculations with the Gogny force, we have subtracted
the kinetic energy of the center of mass motion from the Routhian
to be minimized in order to ensure that the center of mass is kept
at rest. We have also dealt with the exchange Coulomb energy in the
Slater approximation and neglected the contribution of the Coulomb
interaction to the pairing field. For the Gogny force we have used
the parameter set known as D1S that was adjusted more than twenty
years ago \cite{Berger.84,Berger.91} in order to reproduce basic
nuclear matter properties and the binding energies of several magic
nuclei. The HFB wave functions have been expanded in a Harmonic Oscillator
(HO) basis containing 14 major shells which is enough as to grant
convergence in the excitation spectra obtained.

Due to the self-consistent nature of our procedure it is by no means
granted that starting the iterative procedure by blocking the quasi-particle
of lowest energy the minimization process in going to end up in the
lowest energy solution. For this reason one has to repeat the minimization
process several times using different quasi-particle configurations
each time (usually ones with the lowest one quasi-particle energy)
for the initial blocking. In our case we have repeated each calculation
three times implying a computational cost 18 times higher (six values of $J_{z}$
times three starting configurations) than the corresponding
calculation in an even-even neighbor. By following this procedure
we can be pretty sure to have reached the lowest energy solution for all values
of the octupole moment and mass number.

\begin{figure}
\includegraphics[width=0.95\columnwidth]{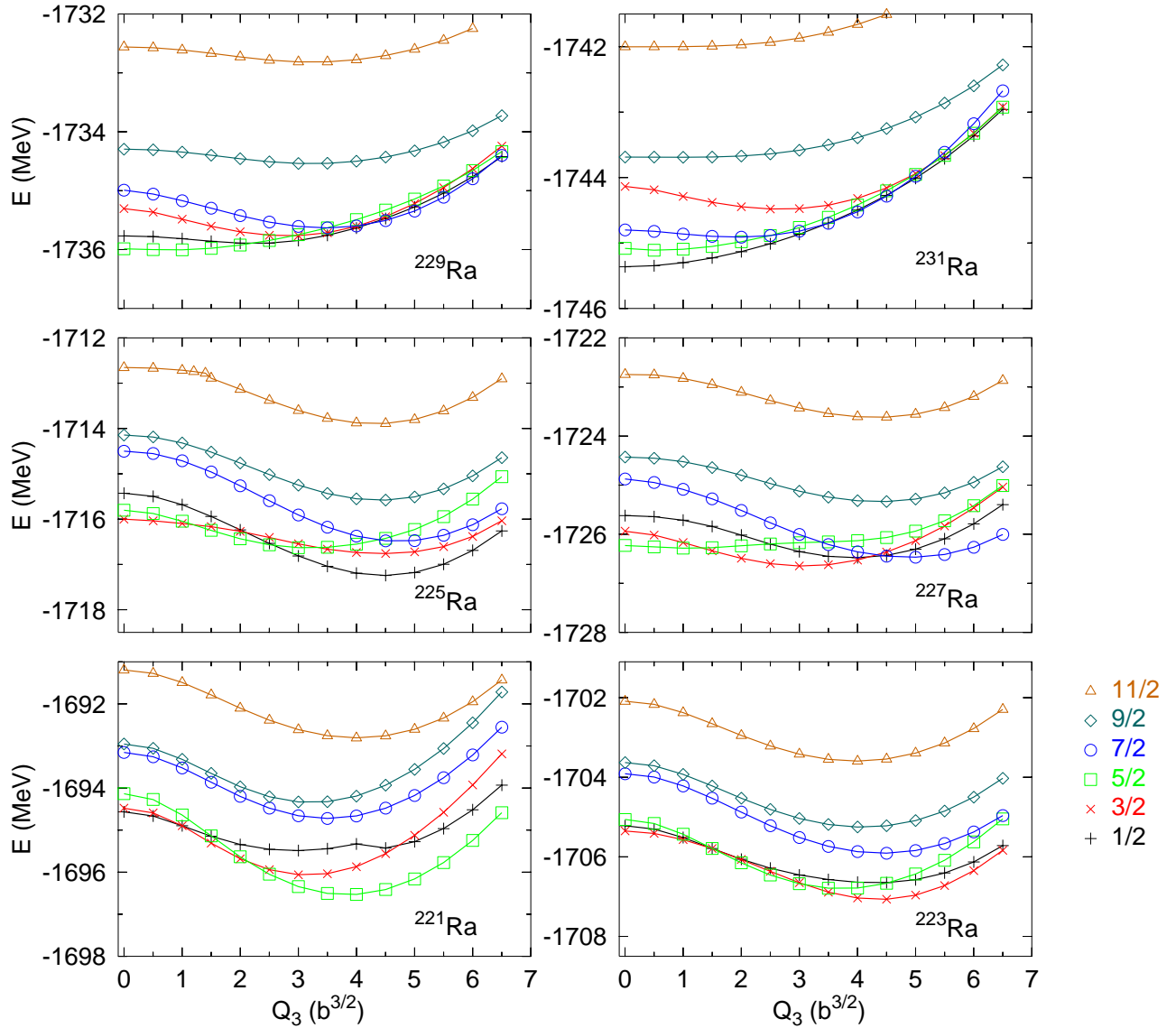}

\caption{(Color online) Potential energy surfaces as a function of the octupole moment $Q_{3}$
(in units of $b^{3/2}\equiv10^{3}\textrm{fm}^{3}$) for the odd-mass
isotopes of Ra considered and blocking in each of the relevant $J_{z}$
channels from $1/2$ till $11/2$. \label{fig:PES}}

\end{figure}

\begin{figure}
\includegraphics[width=0.9\columnwidth]{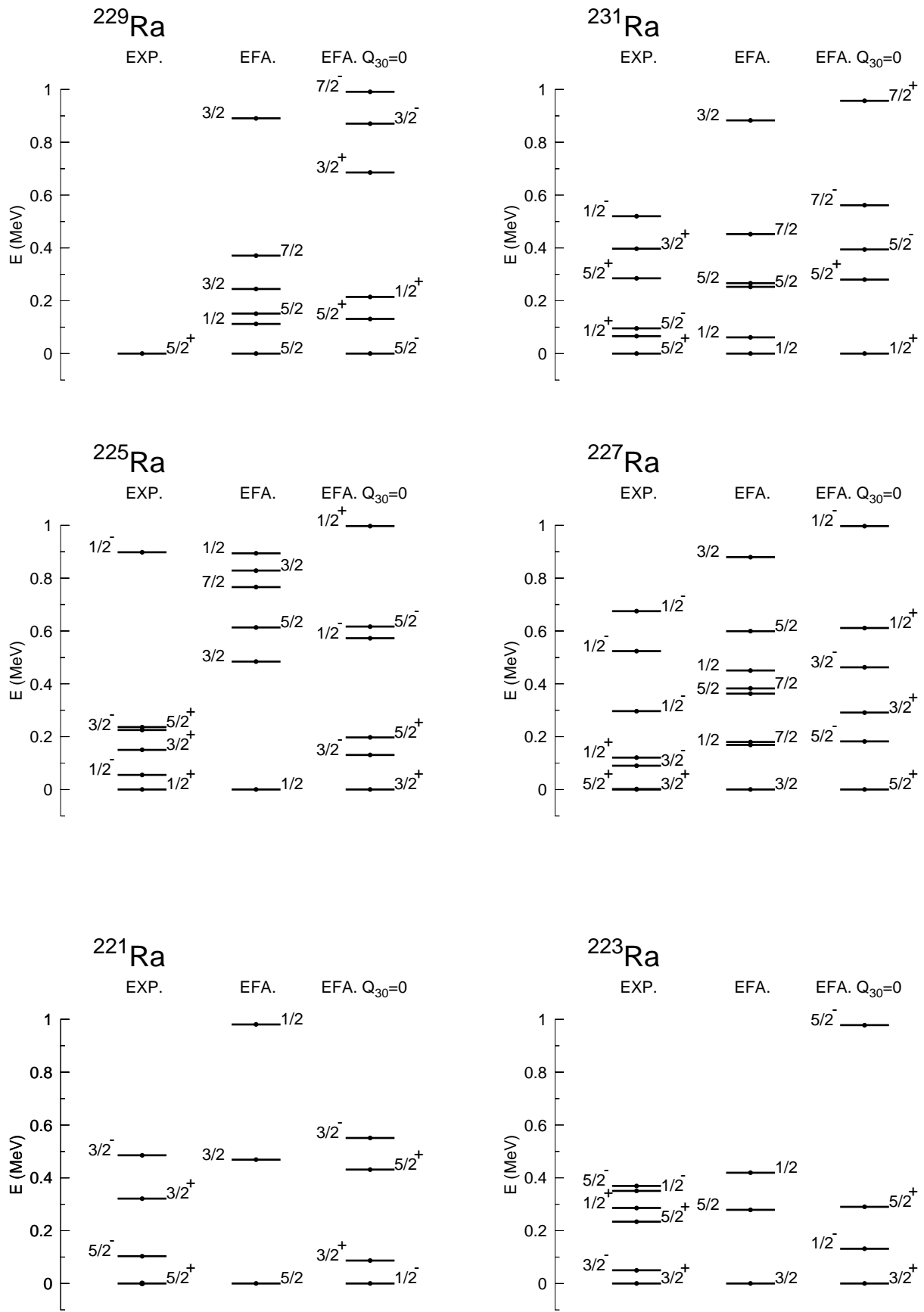}

\caption{Lowest lying excitation spectrum for the six isotopes of Radium considered.
In each panel three spectra are included: the one to the left is the
experimental one, the other two are theoretical predictions (including
octupole deformation effect in the middle and not including it, that
is $Q_{3}=0$, to the right\label{fig:Spectrum})}

\end{figure}

In Fig. \ref{fig:PES} we show the potential energy surfaces (PES)
as a function of the octupole moment for the six Radium isotopes considered
and corresponding to the blocking of the lowest quasi-particles with
$J_{z}$ values ranging from $1/2$ to $11/2$. Higher $J_{z}$ values
are not considered here as the corresponding single particle levels
lie too far away from the Fermi surface as to be relevant for the
lowest energy configurations. By looking at the PES for different
isotopes we learn that the response to octupole deformation 
strongly depends on the $J_{z}$ value of the level blocked as it is for instance
the case in $^{231}$Ra, where the PES of some levels show a minimum
at $Q_{3}=0$ whereas others (like the ones with $J_{z}=3/2$ and $J_{z}=7/2$)
have an octupole deformed minimum at $Q_{3}=3b^{3/2}$. Another interesting
fact is that breaking reflection symmetry implies energy gains of
up to 2 MeV in some cases as compared to the symmetry preserving mean
field configuration and this amount of energy can not be disregarded
in the evaluation of masses. In all the cases, the potential energy
wells are not very deep indicating the relevance of considering fluctuations
on the octupole degree of freedom. This fact was already observed
in calculations of the same kind and the same interaction but for
even-even nuclei \cite{Rob.84,Eggs.89} and the conclusion reached
there was that a treatment of the octupole fluctuations was needed.
One of the advantages of the present formulation of the EFA is the
fact that the standard methods to incorporate correlations (as for
instance the collective Schroedinger equation, see Refs \cite{Rob.84,Eggs.89}
for details) can be now generalized to the present case \cite{Sara.TBP}.
Another improvement to the present treatment is to consider parity
projection in the manner discussed in \cite{Eggs.91} for even-even
nuclei; work along this lines is in progress and will be reported
in the near future \cite{Sara.TBP}.

Another consequence of the different responses to octupole deformation
of the different $J_{z}$ blocked configurations is that the spectrum
corresponding to the minimum energy is rather different from the one
obtained by restricting the system to reflection symmetric configurations
showing the relevance of the octupole degree for freedom for the ordering
of the spectrum of these and other odd-A nuclei in the region of the
Actinide. The different theoretical spectra allowing and not allowing
octupole deformation are depicted in Fig. \ref{fig:Spectrum} along
with the experimental data. The inclusion of octupole deformation
improves the spectrum for several nuclei like $^{221}$Ra, $^{225}$Ra
and $^{227}$Ra and makes it to look much closer to the experimental
one. In fact, the inclusion of octupole deformation on these nuclei
allows for a correct prediction of the spin of the ground state. For
the other cases the inclusion of the octupole degree of freedom leaves
the spectra more or less unchanged as compared to the $Q_{3}=0$
results. On the other hand, it has to be kept in mind that when the
octupole moment is allowed to take values different from zero, parity
mixing is also allowed in the wave functions and therefore the different
levels lose parity as a quantum number. To restore parity symmetry
a projection onto good parity is required that would lead to the appearance
of two levels with opposite parity for each one of the levels breaking
the parity symmetry. The energy splitting between the two levels strongly
depends on the octupole deformation but we can state as a rule of
thumb that the energy splitting is going to be rather small (a few
tens of keV at most) in most of the cases and it will hardly exceed
0.5 MeV. Fortunately, the formalism developed in this paper can be
extended to the situation of symmetry restoration by means of parity
projection and the results as well as the whole formalism will be
published elsewhere. For the present purposes the only relevant information
needed is that parity projection will lead to a parity doublet with
a not so big energy splitting. A bigger splitting could eventually
be obtained by fully treating octupole fluctuations as mentioned above
and again one of the advantages of the present formulation is that
the collective masses needed for such a task can be consistently evaluated
in the present framework. Concerning the comparison with the experiment,
we have to keep in mind the strong sensitivity of the spectra to tiny
details of the underlying single particle states that makes quite
difficult to obtain the experimental spectrum in the right order.
The accuracy of modern effective interaction only entitle to look
for an agreement in the number of levels and $J_{z}$ values in a
range of 1 or 1.5 MeV above the ground state but it does not entitle
whatsoever to sought for an agreement in the ordering of the levels.
However, the inclusion of the octupole degree of freedom allows for
a correct description of the spin of the ground state in five of the
six considered nuclei. With this in mind we can conclude that the
agreement with experiment is quite good in the whole isotopic chain.

A procedure that is sometimes used to describe odd-A nuclei is to 
neglect explicitly all kind of polarization effects and treat the 
quasiparticle excitations in a perturbative fashion \cite{Duguet.01}. 
To this end, a reference HFB wave function $|\varphi_R\rangle$ is 
computed assuming that is fully paired (that is, is a linear combination 
of wave functions with even number of particles) but the number of 
particles is constrained to be odd on the average. The wave functions of the ground state and 
excitations of the odd nucleus are built as one-quasiparticle excitations 
built on top of the reference HFB wave function ($\beta_\mu^ +|\varphi_R\rangle$). 
The excitation energies are then given by the corresponding one-quasiparticle 
energies $E_\mu$ computed as the mean value of the Routhian 
$E_\mu=\langle \varphi_R | \beta_\mu (\hat{H}-\lambda \hat{N})\beta_\mu^+ |\varphi_R\rangle $ 
in an attempt to correct those energies perturbatively for the fact that particle number differs 
from the right value by the quantity 
$N^{11}_{\mu \mu}=\langle\varphi_R| \beta_\mu \hat{N} \beta_\mu^+ |\varphi_R \rangle$. 
The perturbative correction works well when $N^{11}_{\mu \mu}$ is small but it is 
not so reliable when this quantity is large, as the chemical potential is usually a few MeV. 
In Table \ref{Table:Perturbative} we present the results of such perturbative 
calculation for the nuclei $^{223}$Ra and $^{225}$Ra. The spectrum looks rather 
similar to the selfconsistent one obtained in the EFA framework (see Fig. 
\ref{fig:Spectrum}) but the perturbative one is much more compressed. In this
table we also include the values of $N^{11}_{\mu \mu}$ and we observe very big
values of the order of one in absolute value. These large values together with
the neutron chemical potential energies (-5.24 MeV for $^{223}$Ra and -4.97 MeV for 
$^{225}$Ra) make a too big perturbative correction. It has to be mentioned that
the value of $N^{11}_{\mu \mu}$ for the ground state is rather small in agreement
with the motivation  for the introduction of $|\varphi_R\rangle$ given in Ref. \cite{Duguet.01}.
We can conclude that the polarization effects accounted for by the EFA are rather 
strong and the perturbative treatment, although a reasonable qualitative approximation,
is not good at the quantitative level. No other mean values and/or physical quantities
are considered in this comparison as it would imply the evaluation of
the perturbative correction due to particle number departures from the physical values
(that is the evaluation of "chemical potential" like quantities for
mean values of arbitrary observables) and this is out of the scope of the present work.

\begin{table}
\begin{tabular}{|c|c|c||c|c|c|} \hline
\multicolumn{3}{|c|}{$^{223}$Ra} & \multicolumn{3}{|c|}{$^{225}$Ra} \\ \hline
$J_z$ & $E_\mu$ (MeV) & $N^{11}_{\mu \mu}$ & $J_z$ & $E_\mu$ (MeV)  & $N^{11}_{\mu \mu}$ \\ \hline
3/2  & 0.000  &  0.06  & 1/2 & 0.000 & -0.09 \\ \hline
5/2  & 0.059  & -0.56  & 3/2 & 0.211 & -0.62 \\ \hline
1/2  & 0.216  &  0.64  & 5/2 & 0.410 & -0.81 \\ \hline
7/2  & 0.695  &  0.94  & 7/2 & 0.484 &  0.89 \\ \hline
1/2  & 1.077  &  0.93  & 1/2 & 0.573 &  0.89 \\ \hline
\end{tabular}
\caption{Perturbative results for the nuclei $^{223}$Ra and $^{225}$Ra. The lowest five states in each case have
been included.}\label{Table:Perturbative}
\end{table}

\begin{figure}
\includegraphics[width=0.25\columnwidth,angle=270]{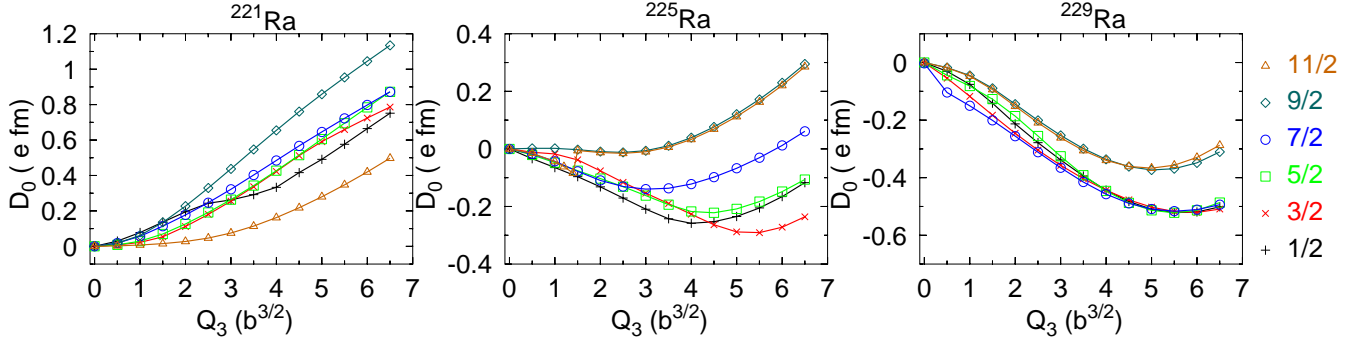}
\caption{(Color online) Dipole moments as a function of $Q_{3}$ for three representative
nuclei and all the $J_{z}$ values  considered\label{fig:Dipole}}
\end{figure}

Another relevant physical quantity for octupole deformed nuclei is
the intrinsic dipole moment $D_{0}$ that is directly related to the
strong $E1$ transition probabilities observed in these nuclei. It
is given as the mean value of the dipole operator
\begin{equation}
D_{0}=e\frac{NZ}{A}(\langle\hat{z}\rangle_{prot}-\langle\hat{z}\rangle_{neut})
\label{eq:Dipole}\end{equation}
in terms of the mean value of the $z$ coordinate for protons and
neutrons. The theoretical results for such quantity and for each blocked
configuration and as a function of $Q_{3}$ are presented in Fig.
\ref{fig:Dipole} for three representative nuclei. In the first
nucleus $^{221}$Ra the global tendency for $D_{0}$ is to increase
with increasing octupole moment for all possible configurations with
different $J_{z}$. On the other hand, for the nucleus $^{225}$Ra
the dipole moment $D_{0}$ first decreases going to negative values
and afterward increases to reach positive values (or negative but
rather small) as the octupole moment increases. Finally, for the nucleus
$^{229}$Ra the dipole moments steadily decrease with increasing octupole
moment and reaching absolute values greater than in the previous case.
The behavior is rather similar to the one of the neighboring even-even
Radium isotopes as can be observed in Ref. \cite{Eggs.89}. This
behavior was related in \cite{Eggs.89} to the increasing occupancy
of the neutron $j_{15/2}$ orbital with increasing number of neutrons
and leads to the prediction of a minimum in the absolute value of
the dipole moment $|D_{0}|$ for the nucleus $ $ $^{224}$Ra. As
a consequence, we expect substantially lower values of $|D_{0}|$
for the nuclei $^{223}$Ra and $^{225}$Ra as is indeed the case. 

The numerical values of the dipole moments obtained at the minima
of the potential energy curves of Fig. \ref{fig:PES} are given
in Table \ref{tab:Dipole}. The values are given not only for the
theoretical ground state but also for other close lying configurations
as our theoretical prediction not necessarily coincides with the experimental
assignments. Experimental values taken from the compilation of \cite{Butler.96}
and also from \cite{Aas.96} tell us that for $^{221}$Ra the $J^{\pi}$
of the ground state is $5/2^{+}$ with a value of $D_{0}=0.36\pm0.10$
e fm which is in good agreement with the $D_{0}$ value of the lowest
lying $5/2^{+}$ theoretical state. For the $^{223}$Ra nucleus there
are values not only for the ground state but also for some excited
ones; the values of $D_{0}$ are $0.129\pm0.009$ e fm, $0.035\pm0.005$
e fm and $0.076\pm0.004$ e fm for the $3/2^{+}$(ground state), $5/2^{+}$
and $1/2^{+}$states, respectively. As it can be checked in Table
\ref{tab:Dipole} the agreement between theory and experiment is
very satisfactory for the three states. In the nucleus $^{225}$Ra
the experimental $D_{0}$ value is $0.14\pm0.02$ e fm that corresponds
in reality to the absolute value of that quantity ($D_{0}$ is extracted
from $B(E1)$ values where it enters squared and therefore the sign
can not be determined in that way). Taking this in account, we can
say that there is a reasonable agreements with the theoretical prediction
which, on the other hand, can be quite strongly affected by fluctuations
in the octupole degree of freedom \cite{Eggs.89}. Finally, for the
$^{227}$Ra isotope the experimental value is $0.099\pm0.003$ e fm
for the $3/2^{+}$ ground state and this value again agrees reasonably
well with the theoretical prediction. The agreement between theory
and experiment can be considered as quite good specially taking into
account the fact that no information on this kind of physics was included
in the fitting procedure of the force afterwards.

\begin{table}
\begin{tabular}{|c|c||c||c||c||c||c|}
\hline 
$J_{z}$ & $^{221}$Ra & $^{223}$Ra & $^{225}$Ra & $^{227}$Ra & $^{229}$Ra & $^{231}$Ra\tabularnewline
\hline
\hline 
$\frac{1}{2}$ & 0.331 &  0.056  & -0.253 & -0.413 & -0.226 & 0.000\tabularnewline
\hline 
$\frac{3}{2}$ & 0.255 & 0.175 & -0.263 & -0.275 & -0.357 & -0.374\tabularnewline
\hline 
$\frac{5}{2}$ &  0.427 & 0.050 & -0.162 & -0.053  & -0.077  & -0.057\tabularnewline
\hline 
$\frac{7}{2}$ & 0.403 & 0.254  & -0.098 &  -0.471 & -0.415 & -0.279\tabularnewline
\hline
\end{tabular}

\caption{Dipole moments $D_{0}$ in e fm for the six nuclei considered and obtained
for the configuration corresponding to the minimum of each blocked
configuration with varying $J_{z}$ value. Only the $J_{z}$ values
up to 7/2 have been considered as this is the maximum value of that
quantity for all the low lying states.\label{tab:Dipole}}

\end{table}

\section{Conclusions}

A prescription for the treatment of odd mass nuclei in a time reversal
preserving mean field (HFB) framework usually known as the Equal Filling
Approximation has been justified in terms of standard procedures of
quantum statistical mechanics. It turns out that the EFA can be described
as a mixed state where the blocked one-quasi-particle state and its
time-reversed counterpart have probability one whereas the others
have zero probability. As a consequence, the EFA energy is given by
an average involving the energy of the underlying even-even system,
the energy of the blocked one-quasi-particle configurations and the
two quasi-particle excitation built out of them. As the energy now
has a well defined expression in terms of the HFB wave functions it
is possible to invoke the variational principle to obtain the standard
EFA-HFB equation and allowing for the use of more sophisticated numerical
techniques, like the gradient method, for its numerical solution.
The method has been applied to the study of odd-A Radium isotopes
as a function of octupole deformation and with the Gogny D1S force
and the agreement obtained between theory and experiment is quite
reasonable. One of the advantages of the present method is the preservation
of time-reversal symmetry that reduces substantially the computational
cost of mean field calculations of odd-mass nuclei. Another advantage
of the justification obtained in this paper is that the procedure
can be extended beyond mean field in a consistent way increasing its
range of applicability.

\begin{acknowledgments}
This work was supported in part by DGI, Ministerio de Ciencia y Tecnolog\'\i a,
Spain, under Project FIS2004-06697. S. P-M. acknowledges financial support from
the Programa de Formaci\'on del Profesorado Universitario (Ref. AP
2001-0182).
\end{acknowledgments}

\appendix

\section{Higher order degeneracy: the spherical case.}

In the spherical case, the levels to be {}``blocked'' are characterized
by angular momentum quantum numbers $j_{B},m_{B}$ with $m_{B}=-j_{B},\ldots,j_{B}$.
In the spirit of the EFA each of these levels will be uniformly populated with
a fraction $1/(2j_{B}+1)$ of a nucleon what means that in this case
the density matrix and pairing tensors in the EFA approximation are
given by the {}``average'' density 
\begin{equation}
\rho_{kk'}^{EFA}=\left(V^{*}V^{T}\right)_{kk'}+
\frac{1}{2j_{B}+1}\sum_{m_{B}=-j_{B}}^{j_{B}}
\left(U_{k'\mu_{B},j_{B},m_{B}}
U_{k\mu_{B},j_{B},m_{B}}^{*}-
V_{k'\mu_{B},j_{B},m_{B}}^{*}V_{k\mu_{B},j_{B},m_{B}}\right)\label{eq:}
\end{equation}
and the {}``average'' pairing tensor 
\begin{equation}
\kappa_{kk'}^{EFA}=\left(V^{*}U^{T}\right)_{kk'}+
\frac{1}{2j_{B}+1}\sum_{m_{B}=
-j_{B}}^{j_{B}}\left(U_{k\mu_{B},j_{B},m_{B}}
V_{k'\mu_{B},j_{B},m_{B}}^{*}-
U_{k'\mu_{B},j_{B},m_{B}}V_{k\mu_{B},j_{B},m_{B}}^{*}\right)\label{eq:}
\end{equation}
The EFA occupancies to be used in the density matrix operator are
in this case
\begin{equation}
f_{\mu}=\left\{ \begin{array}{ccc}
\frac{1}{2j_{B}+1} &  & \mu=\mu_{B},j_{B},m_{B}\;\;\; m_{B}=-j_{B},\ldots,j_{B}\\
0 &  & \textrm{otherwise}\end{array}\right.\label{eq:}
\end{equation}
where the index $\mu$ has been decomposed in the labels $j_{B}$
corresponding to the total angular momentum, $m_{B}$ corresponding
to the third component of the angular momentum and finally $\mu_{B}$
which represents the remaining quantum numbers. Once the value of
the EFA occupancies are established the formalism developed in the
main body of the paper can be applied straightforwardly and all the
formulas can be used verbatim.

\section{Variational parameters and gradients}

Given a reference HFB wave function $|\phi\rangle$ the most general
HFB wave function $|\phi(\mathbb{Z})\rangle$, not orthogonal to it,
is given by $|\phi(\mathbb{Z})\rangle=\exp(i\hat{Z})|\phi\rangle$
where $\hat{Z}$ is an hermitian (to preserve the unitarity of the
transformation) one-body operator $\hat{Z}=\frac{1}{2}\sum_{\mu\nu}\mathbb{Z}_{\mu\nu}\alpha_{\mu}^{+}\alpha_{\nu}$
which is written in terms of the generalized quasi-particle operators
$\alpha_{\mu}=(\beta_{1},\ldots,\beta_{N},\beta_{1}^{+},\ldots,\beta_{N}^{+})$,
its hermitian conjugate $\alpha_{\mu}^{+}$ and the bipartite hermitian
matrix
\begin{equation}
\mathbb{Z}=\left(\begin{array}{cc}
Z^{11} & Z^{20}\\
-Z^{20*} & -Z^{11*}\end{array}\right)\label{eq:}
\end{equation}
that parametrizes the Bogoliubov transformation. Not all the $2\times(2N)^{2}$
parameters of this matrix are independent as the matrix $Z_{mn}^{11}$
has to be an hermitian matrix (with $N^{2}$ free parameters, the
complex numbers $Z_{mn}^{11}$ with $m>n$ plus real $Z_{mm}^{11}$,
i.e. $N^{2}=2\times\left(N(N-1)/2\right)+N$ ) whereas $Z^{20}$ is
a complex skew-symmetric matrix (with $N^{2}-N$ free parameters,
the complex numbers $Z_{mn}^{20}$ with $m>n$, i.e. $N^{2}-N=2\times\left(N(N-1)/2\right)$
). As customary we will consider $Z^{11}$ and $Z^{11\,*}$ as independent
parameters instead of the real and imaginary parts of $Z^{11}$ and
will apply the same consideration to $Z^{20}$. As a consequence,
the variational parameters of the Bogoliubov transformation are $Z_{mn}^{11}$
and $Z_{mn}^{11\:*}$ with $m>n$, the real parameters $Z_{mm}^{11}$
and finally $Z_{mn}^{20}$ and $Z_{mn}^{20\,*}$ with $m>n$. The
variational parameters can be handled in a compact notation by introducing
the vector $z_{\rho}$ of dimension $2N^{2}-N$
\begin{equation}
z_{\rho}=\left.\left\{ \begin{array}{ccc}
Z_{mn}^{11} & m>n\\
Z_{mm}^{11}\\
Z_{mn}^{11\,*} & m>n\\
Z_{mn}^{20} & m>n\\
Z_{mn}^{20\,*} & m>n\end{array}\right.\right.\label{eq:}
\end{equation}
As  obtained in the body of the paper, the variation of the
mean value of an observable can be written as 
\begin{equation}
\delta a=\frac{i}{2}\mathrm{Tr}_{2}[\mathbb{O}\mathbb{Z}]+
O(\mathbb{Z}^{2})\label{eq:}
\end{equation}
where $\mathbb{O}=[\mathbb{R},\mathbb{A}]$. Taking the most general
form of the bipartite matrix 
\begin{equation}
\mathbb{O}=\left(\begin{array}{cc}
O^{11} & O^{12}\\
O^{21} & O^{22}\end{array}\right)\label{eq:}
\end{equation}
a little algebra gives $\delta a$ as a function of the variational
parameters \begin{equation}
\delta a=\frac{i}{2}\left(\sum_{m>n}\left(O_{nm}^{11}-O_{mn}^{22}\right)Z_{mn}^{11}+\left(O_{mn}^{11}-O_{nm}^{22}\right)Z_{mn}^{11*}+\left(O_{nm}^{21}-O_{mn}^{21}\right)Z_{mn}^{20}+\left(O_{mn}^{12}-O_{nm}^{12}\right)Z_{mn}^{20*}+\sum_{m}\left(O_{mm}^{11}-O_{mm}^{22}\right)Z_{mm}^{11}\right)\label{eq:}\end{equation}
 Taking now into account the expression of $\mathbb{O}$ and the fact
that $\mathbb{A}$ is the quasi-particle representation of the operator
given by
\begin{equation}
\mathbb{A}=\left(\begin{array}{cc}
A^{11} & A^{20}\\
-A^{20*} & -A^{11*}\end{array}\right)\label{eq:}
\end{equation}
where $A^{20}$ is a skew-symmetric and $A^{11}$ is a hermitian matrix
if the operator is hermitian (as it should be for any observable !)
we obtain
\begin{equation}
\delta a=i\sum_{m>n}A_{nm}^{11}(f_{n}-f_{m})
Z_{mn}^{11}-A_{nm}^{20*}(1-f_{n}-f_{m})
Z_{mn}^{20}+\mathrm{c.c.}\label{eq:}
\end{equation}
This expression can be written in a compact way $\delta a=\sum_{\rho}z_{\rho}(g_{a})_{\rho}$
by introducing the vector $\left(g_{a}\right)_{\rho}$ which is the
gradient of the mean value $a$ with respect to the variational parameters
\begin{equation}
\left(g_{a}\right)_{\rho}=\left.\left\{ \begin{array}{ccc}
iA_{nm}^{11}(f_{n}-f_{m}) & m>n\\
0 & m=m\\
-iA_{nm}^{11*}(f_{n}-f_{m}) & m>n\\
-iA_{nm}^{20*}(1-f_{n}-f_{m}) & m>n\\
iA_{nm}^{20}(1-f_{n}-f_{m}) & m>n\end{array}\right.\right.\label{eq:}
\end{equation}
Finally, it is important to point out that the requirement 
$\left(g_{a}\right)_{\rho}=0$
is equivalent to $[\mathbb{R},\mathbb{A}]=0$, a fact that is
used in the derivation of the EFA-HFB equation.

\end{document}